\documentclass[conference]{IEEEtran}
\IEEEoverridecommandlockouts
\usepackage{cite}
\usepackage{amsmath,amssymb,amsfonts}
\usepackage{algorithmic}
\usepackage{graphicx}
\usepackage{textcomp}
\usepackage{amsmath}
\usepackage{xcolor}
\usepackage{multirow}
\usepackage{array}
 \usepackage{url}
  \usepackage[colorlinks]{hyperref}
 \usepackage{orcidlink}

\def\BibTeX{{\rm B\kern-.05em{\sc i\kern-.025em b}\kern-.08em
    T\kern-.1667em\lower.7ex\hbox{E}\kern-.125emX}}
\begin{document}

\title{Conference Paper Title*\\
{\footnotesize \textsuperscript{*}Note: Sub-titles are not captured in Xplore and
should not be used}
\thanks{Identify applicable funding agency here. If none, delete this.}
}
\title{DATCloud: A Model-Driven Framework for Multi-Layered Data-Intensive Architectures
\\}

\author{
    \IEEEauthorblockN{Moamin Abughazala\IEEEauthorrefmark{1}\IEEEauthorrefmark{2}, Henry Muccini\IEEEauthorrefmark{1}} \\
    \IEEEauthorblockA{\IEEEauthorrefmark{1} DISIM Department, University of L’Aquila, L’Aquila, Italy \\
    \{moamin.abughazala1, henry.muccini\}@univaq.it} \\[5pt]
    \IEEEauthorblockA{\IEEEauthorrefmark{2} Department of Information Technology, An Najah N. University, Nablus, Palestine \\
    m.abughazaleh@najah.edu}
}




\maketitle
\begin{abstract}

The complexity of multi-layered, data-intensive systems demands frameworks that ensure flexibility, scalability, and efficiency. DATCloud is a model-driven framework designed to facilitate the modeling, validation, and refinement of multi-layered architectures, addressing scalability, modularity, and real-world requirements. By adhering to ISO/IEC/IEEE 42010 standards, DATCloud leverages structural and behavioral meta-models and graphical domain-specific languages (DSLs) to enhance reusability and stakeholder communication. Initial validation through the VASARI system at the Uffizi Gallery demonstrates a 40\% reduction in modeling time and a 32\% improvement in flexibility compared to manual methods. While effective, DATCloud is a work in progress, with plans to integrate advanced code generation, simulation tools, and domain-specific extensions to further enhance its capabilities for applications in healthcare, smart cities, and other data-intensive domains.
\end{abstract}

\section{Introduction} 
The increasing complexity of data-intensive systems, driven by the rapid expansion of Internet of Things (IoT) devices, has introduced significant challenges for system modeling. Modern architectures must address scalability, modularity, and data management across multi-layered cloud, fog, and edge systems \cite{8364039}. However, existing frameworks often lack robust modeling capabilities, resulting in inefficiencies when designing, adapting, and managing these architectures \cite{bauer2013iot}.

To address these challenges, this paper introduces \textbf{DATCloud} a model-driven framework designed to simplify and enhance the modeling of multi-layered, data-intensive systems. 
Unlike traditional methods DATCloud models and combines data systems across cloud, fog, and edge layers. 
DATCloud uses structural and behavioral meta-models to represent the logical architecture of systems, including components, data flows, and workflows. These meta-models provide architects with the tools needed to define, validate, and refine system designs efficiently. The framework further incorporates graphical domain-specific languages (DSLs), offering an intuitive interface for creating models that ensure consistency and compliance with architectural standards such as ISO/IEC/IEEE 42010 \cite{42010}.

A key focus of DATCloud is improving the efficiency and flexibility of the modeling process. By supporting templates,and modular design, the framework reduces the effort required for iterative development and simplifies communication between stakeholders. These capabilities enable architects to handle complex architectures while minimizing manual overhead.


Research, such as Monitor-IoT \cite{9791260}, SimulatIoT \cite{9465106}, Silva et al.\cite{SILVA20132005}, and CHESSIoT \cite{ihirwe2024chessiot}, has highlighted the potential of multilayer monitoring architectures for IoT systems. These studies cover software, hardware, physical, and deployment aspects, but reveal a notable gap in the \textit{Data View}.

The rest of this paper is structured as follows: Section II reviews related work, highlighting existing frameworks and their limitations in addressing the challenges of modeling multi-layered, data-intensive systems. Section III describes the design of DATCloud, detailing its structural and behavioral meta-models, core functionalities, and adherence to ISO/IEC/IEEE 42010 standards. Section IV presents the case study and discussion of initial results for DATCloud, focusing on metrics for modeling time and flexibility. It highlights stakeholder insights and demonstrates DATCloud's practical effectiveness in managing visitor flow and architectural flexibility through its application in the VASARI system at the Uffizi Gallery. Finally, Section V outlines future directions for enhancing the framework, including advanced code generation and domain-specific extensions, and concludes with a summary of its contributions.

\section{Related Work}
The rapid growth of IoT systems and data-intensive applications has prompted the development of various frameworks and methodologies for modeling multi-layered architectures. While significant progress has been made, some gaps still exist, particularly in the \textbf{data view} and in \textbf{integrating data-intensive applications across the cloud, fog, and edge layers}. This section reviews the key contributions in these areas and highlights the gaps that DATCloud aims to address.

\subsection{Modeling IoT Multi-Layer Architectures}

Efficient architectures are essential for IoT systems to process distributed data across edge, fog, and cloud layers. Various frameworks have been proposed to model and manage these complex systems effectively.

\begin{itemize}
    \item Bauer et al. \cite{bauer2013iot} introduced an IoT Reference Architecture emphasizing the roles of cloud, fog, and edge layers. While this model outlines task allocation, it lacks a concrete framework for modeling workflows and data interactions between these layers.    
    \item Taivalsaari and Mikkonen \cite{taivalsaari2018development} focus on IoT systems' lifecycle, emphasizing the interplay between edge devices and the cloud. However, their approach primarily addresses operational challenges rather than architectural modeling.
    
    \item Ihirwe et al. \cite{ihirwe2024chessiot} proposed CHESSIoT, a model-driven engineering (MDE) framework for IoT system design. This framework leverages domain-specific languages (DSLs) to model and verify IoT architectures. While effective for IoT-specific use cases, CHESSIoT does not generalize to multi-layered systems involving complex data workflows.

\end{itemize}

While current frameworks recognize the hierarchical structure of IoT systems, they mainly concentrate on system behavior or task allocation. They do not adequately address the \textbf{data view}, which creates a gap in \textit{how data is modeled, stored, and processed across multi-layered architectures}. Additionally, most frameworks are specific to certain domains and lack the flexibility needed for wider applications.

\subsection{Modeling Data-Intensive Applications}
Data-intensive applications are defined by their need to manage, process, and analyze large volumes of data across distributed systems. Effective modeling of such systems requires frameworks that prioritize data workflows, scalability, and adaptability.

\begin{itemize}
    \item Kleppmann \cite{kleppmann2017designing} explores the challenges of building scalable, distributed systems for handling data pipelines and real-time analytics. While this work provides a thorough understanding of data-intensive systems, it focuses on implementation rather than architectural modeling.

    \item Abughazala et al. \cite{abughazala2023architecture} \cite{abughazala2022dat} \cite{10092710} proposed a Framework, one of the few works explicitly addressing modeling data-intensive workflows. The Framework introduces meta-models to describe how data flows between components. Its focus on industrial data workflows limits its relevance to general multi-layered architectures, including cloud, fog, and edge systems.

    \item Raj \cite{raj2020modelling} introduced a conceptual model for designing a data pipeline consisting of two primary components: nodes and connectors. The nodes serve as the core abstract data units, while the connectors facilitate data transmission and communication between these nodes.

    \item Borelli \cite{borelli2020architectural} proposed a classification framework for key software components and their interrelationships to model software architectures tailored for specific IoT applications. These components are presented as abstract representations.

\item Erraissi \cite{erraissi2018data, erraissi2019big} developed a meta-model encompassing data sources, ingestion layers, and a Big Data visualization layer to provide a structured approach to managing and visualizing Big Data workflows.

\end{itemize}

Current frameworks for data-intensive applications \textbf{lack a unified approach to modeling data workflows across cloud, fog, and edge layers}. Instead of providing holistic integration, they often focus on isolated aspects. This gap highlights the need for a scalable and modular framework, such as DATCloud, to effectively tackle these complexities and ensure practical applicability in real-world scenarios.
\section{Framework Design}
\label{sec:Methodology}
This section introduces DATCloud’s framework design, focusing on two key components: the \textbf{structural meta-model}, which defines the architecture of multi-layered systems, and the \textbf{behavioral meta-model}, which captures workflows and interactions. Together with graphical DSLs, these meta-models enable scalable, modular, and standards-compliant modeling for data-intensive systems. 

\subsection{Structural Meta-Model}
\begin{center}
  \begin{figure*}[!h]
	\centering
	\makebox[\textwidth]
	{
	    \includegraphics[width=0.87\paperwidth]
	    {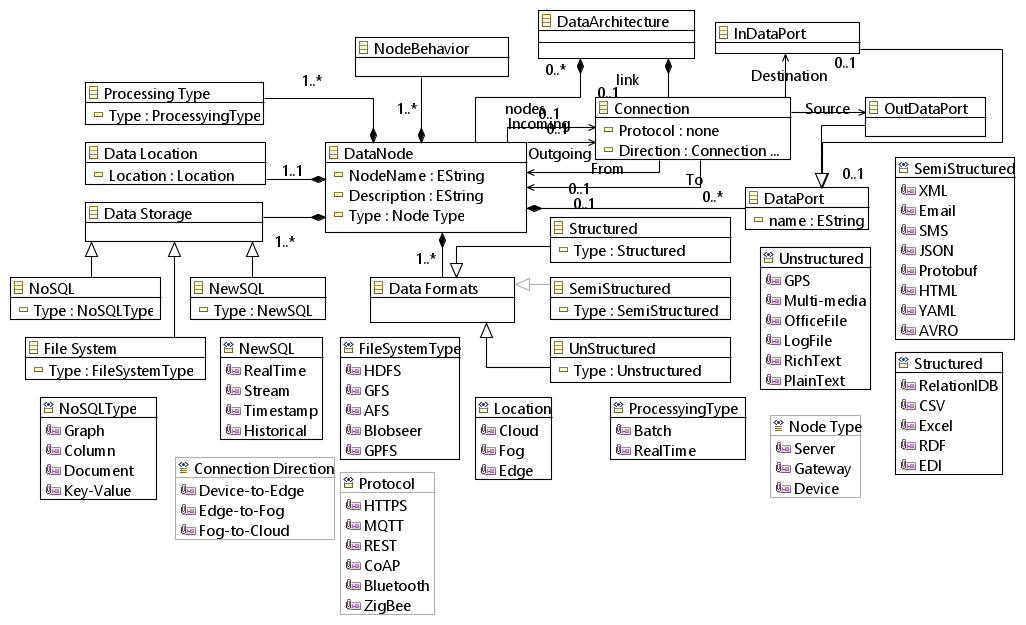}
	}
	\caption{Meta-model: structural concepts}
	\label{fig:dv_mm_s}
    \end{figure*}
\end{center}

Using the structural meta-model, users can establish architecture across cloud, fog, and edge layers, model nodes and connections, choose data storage types like NoSQL or NewSQL, and select communication protocols such as MQTT or REST. They can also illustrate data formats and data flow.
Figure \ref{fig:dv_mm_s} illustrates the structural concepts of the Data Architecture Modeling Language (DAML) meta-model, showcasing the primary components and their interrelations within data-intensive architectures. The DAML structural meta-model defines the physical and logical structure of data systems, emphasizing their nodes, formats, storage, and connections. The key elements are:

\label{sec:meta-model}

\begin{itemize}
    \item \textit{Data Architecture}: Represents the overarching framework that integrates multiple DataNodes and their interactions, encapsulating the architecture of the data system.

    \item \textit{DataNode}: The central element of the model, representing a computational or storage unit responsible for handling data operations \cite{component}. Attributes include:
    \begin{itemize}
        \item NodeName: The identifier of the DataNode.
        \item Description: A description of the DataNode’s role.
        \item Type: Specifies the kind of node, such as Server, Gateway, or Device.
    \end{itemize}
    
    \item \textit{ProcessingType}: Categorizes the nature of data handling within a DataNode, such as Batch processing (Processes data in large, pre-defined batches), Real-Time processing (Processes data continuously as it streams).
    \item \textit{Data Location}: Defines the physical or virtual storage location of the data, which could be Cloud, Fog, or Edge systems.

    \item \textit{Data Storage}: Describes the storage solutions utilized in the architecture, classified as:
    \begin{itemize}
        \item NoSQL: Includes databases such as Graph, Column, Document, and Key-Value stores.
        \item NewSQL: Represents scalable, modern relational databases that support real-time, historical, or stream data.
        \item File System: Includes distributed and scalable storage types like HDFS, GFS, Blobseer, and GPFS.
    \end{itemize}
    \item \textit{Data Formats}: Specifies the types of data handled by the DataNode, categorized into:

Structured: Relational databases, CSV, Excel, RDF, EDI.
Semi-Structured: XML, Email, JSON, YAML, AVRO, and Protobuf.
Unstructured: GPS data, multimedia files, logs, office files, and plain text.
    \item \textit{DataPort}: Represents the interfaces (input/output ports) used for data exchange between nodes:

InDataPort: For receiving data.
OutDataPort: For sending processed data.
    \item \textit{Connection}: Describes the links between DataPorts, enabling data transfer within the system. These are categorized by direction (Incoming and Outgoing) and protocol (e.g., HTTPS, MQTT, REST).

    \item \textit{NodeBehavior}: Specifies the functional operations performed by a DataNode, including its responses to actions and events.

    \item \textit{Incoming/Outgoing Connections}: Define the flow direction of data within the architecture, crucial for understanding dependencies and workflows.
\end{itemize}
This structural meta-model provides a clear and detailed representation of the components, storage methods, data formats, and interactions within a data-intensive system, forming the basis for cloud data architecture modeling.

\subsection{Behavioral Meta-model}

DATCloud allows end-users to define system component behavior using a graphical DSL. The behavioral meta-model supports workflows for:
Data Ingestion, Data Processing, Data Output.
Figure \ref{fig:dv_mm_b} depicts the Behavioral Meta-model of the Cloud Data Architecture Modeling Language (Cloud-DAML), focusing on the internal activities and workflows within DataNodes. This meta-model outlines how data is processed, analyzed, and consumed, detailing the roles of actions, events, and interactions. The key components are:

\begin{center}
 \begin{figure*}[!ht]
	\centering
	\makebox[\textwidth]
	{   	\includegraphics[width=0.87\paperwidth]{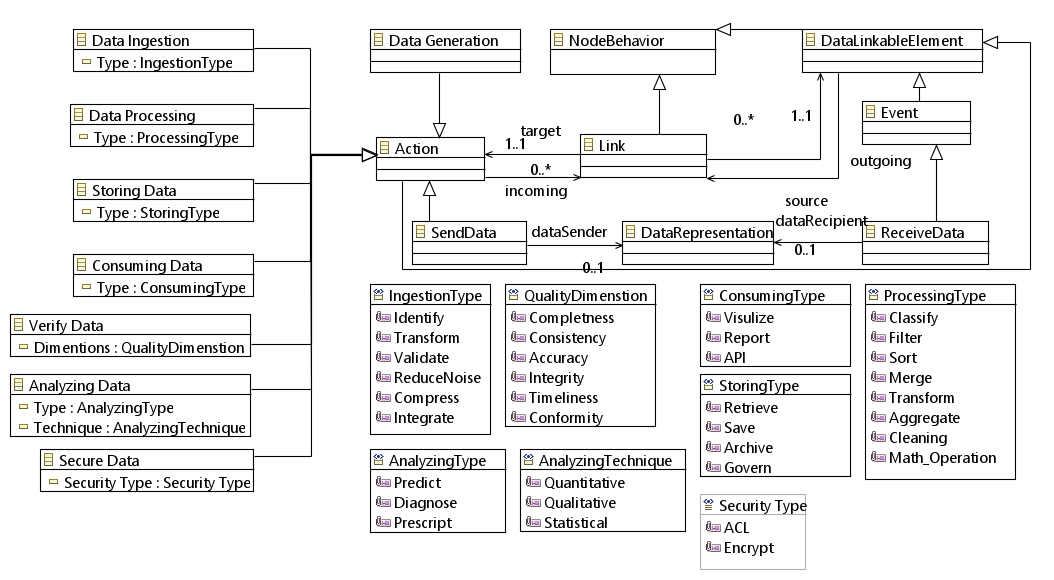}
	}
	\caption{Meta-model: behavioral concepts}
	\label{fig:dv_mm_b}
    \end{figure*}
\end{center}

\begin{itemize}
    \item \textit{NodeBehavior}: Represents the internal behavior of a DataNode, encapsulating the logic, rules, and sequences governing data handling.

    \item \textit{Elements}: These include actions and events that define the step-by-step data processing tasks and their triggers.
    \item \textit{DataPorts}: 
    \begin{itemize}
        \item InDataPort: Serves as an entry point for data into a node.
        \item OutDataPort: Functions as an exit point for processed data.
    \end{itemize}
    
    \item \textit{Actions}:
Core behavioral elements representing atomic tasks executed within a node. Actions can be triggered by events or due to previous steps in the workflow. Examples include:
    \begin{itemize}
        \item Data Generation: Creation or sourcing of new data.
        \item Data Ingestion: Data transfer to a staging area for processing.
        \item Data Processing: Transformations and computations (e.g., filtering, aggregation, analysis).
        \item Storing Data: Saving processed data to databases or data lakes.
        \item Analyzing Data: Performing in-depth analysis to generate insights.
        \item Consuming Data: Utilizing processed data for visualization, reporting, or APIs.
        \item Verifying Data: Ensuring data quality by checking completeness, accuracy, and consistency.
        \item Secure Data: Applying security measures such as ACLs (Access Control List) and encryption to protect data during processing and transfer.
    \end{itemize}
\item \textit{Events}:

Triggered by external stimuli or preceding actions within the system. Example:
 \textit{ReceiveData}: Handles incoming data via an InDataPort, initiating the processing workflow.
\item \textit{Connections}: Links: Define logical pathways and dependencies between actions and events, specifying the sequence of operations and data flow.

\end{itemize}

This behavioral meta-model complements the structural model by detailing the internal workflows and interactions of DataNodes, providing a comprehensive framework for modeling data lifecycle within cloud Data Architecture.

\subsection{Adherence to ISO/IEC/IEEE 42010 Standards} 
In accordance with the ISO/IEC/IEEE 42010 standard \cite{42010}, DATCloud offers multiple architectural views and system components are effectively described by their workflows through structural and behavioral models. Their structural and behavioral meta-models are used to guarantee the clear representations of both the logical and physical architectures and are consistent and complete. To improve stakeholder communication, DATCloud has adopted the use of graphical domain specific languages (DSLs) that generate easily understandable visual models to support the interface between technical and non-technical stakeholders. It also has automated validation to ensure that models are coherent, exhaustive and that they meet stakeholder's needs. Its modular and scalable meta-models enable it to be adaptable for iterative development and changing needs.

\subsection{How To Use DATCloud}
DATCloud defines the architecture using the structural meta-model, specifying nodes, data flows, storage types, and communication protocols. The behavioral meta-model captures workflows within and between nodes, such as data ingestion and processing. Pre-defined templates and reusable components streamline the process, while automated tools validate the model for consistency and resolve dependencies. Iterative refinements ensure alignment with system requirements. DATCloud supports scalability, enabling quick updates to workflows or architecture, and minimizes manual effort through its intuitive interface and automation.

\section{DISCUSSION OF INITIAL RESULTS} 
\label{sec:evalSec}
This section examines DATCloud's application to the VASARI system at the Uffizi Gallery, validating its effectiveness in real-world scenarios. Metrics such as modeling time and flexibility were evaluated using participant logs and feedback. The results highlight DATCloud’s ability to streamline workflows, adapt to changes, and address challenges in multi-layered, data-intensive architectures.
\subsection{CASE STUDY}
\begin{center}
 \begin{figure*}[!th]
	\centering
	\makebox[\textwidth]
	{   	\includegraphics[width=0.89\paperwidth]{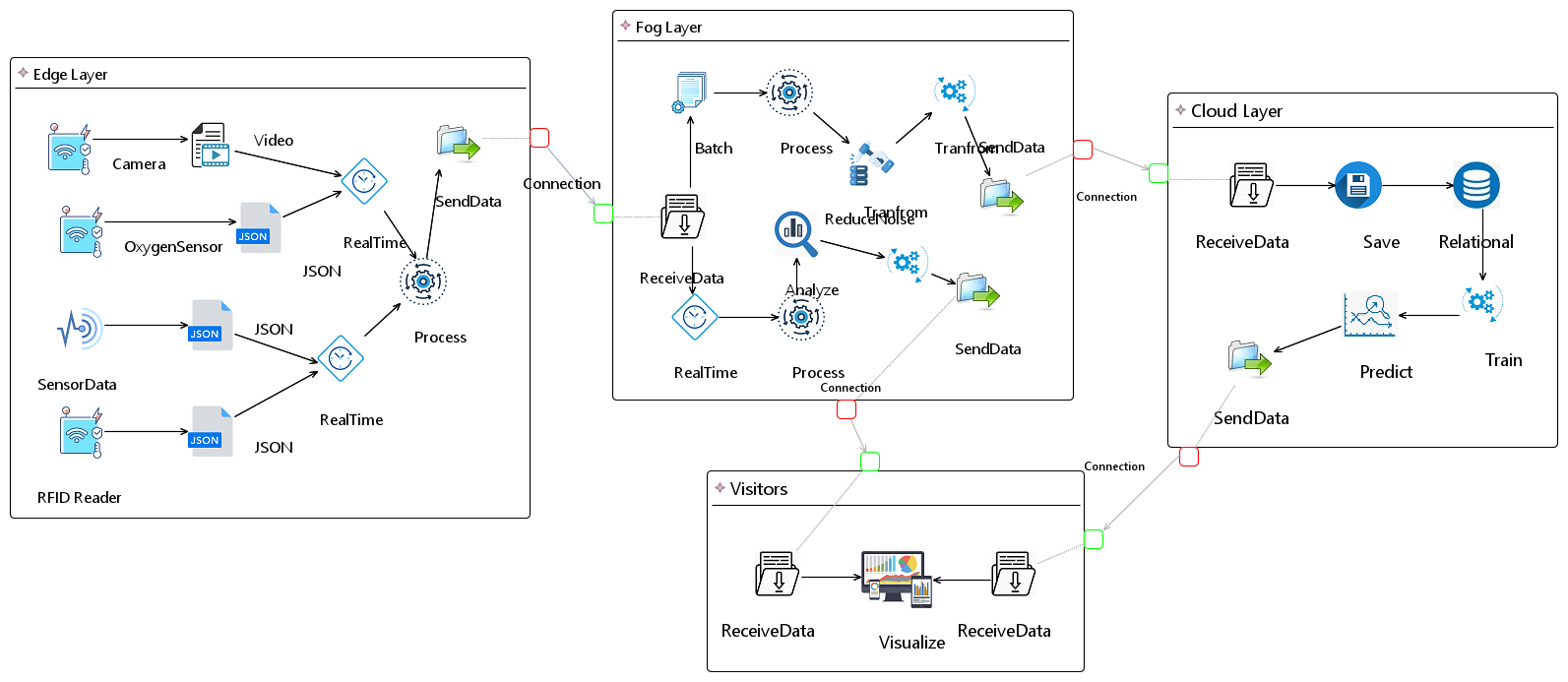}
	}
	\caption{A DATCloud Application for Uffizi Gallery}
	\label{fig:app}
    \end{figure*}
\end{center}

The VASARI system has been deployed at the Uffizi Gallery to control the flow of visitors and has been used to validate the effectiveness of DATCloud. This case study is a model of architecture with IoT sensors at the edge layer, fog nodes for processing, and cloud for analytics. The structural meta-model of the DATCloud modeled the relations of these components, and the behavioral meta-model defined the workflows for visitor tracking and queue management. Figure \ref{fig:app} illustrates how DATCloud models the structure of the Uffizi Gallery. It shows the IoT sensors, the data processing at the fog layer, and the centralized cloud analytics platform. The diagram highlights the flow of visitor data from the sensors to analytics, showcasing DATCloud's flexibility and modularity in facilitating these processes.

\subsection{Initial Results}

To assess the performance of DATCloud in the VASARI system, two key metrics were evaluated: Modeling Time and Flexibility in Architectural Design. Table \ref{tab:time} demonstrates the significant reduction in modeling time achieved using DATCloud, while Table \ref{table:flex} outlines the framework's ability to improve flexibility when adapting to new requirements. For example, the use of pre-defined templates reduced workflow definition time by 50\%, while modular design allowed for faster addition of architectural layers.
\subsection{Metrics and Methodology}
The results in Tables \ref{tab:time} and \ref{table:flex} were derived from structured time logs maintained by the participants during the case study. Participants included system architects, data analysts, and museum staff responsible for managing the VASARI system. Each participant logged their time spent on specific modeling tasks, such as defining workflows, validating the system, and refining models, using both manual methods and DATCloud. The average values for modeling time ($T_{\text{base}}$ and $T_{\text{DAT}}$) and flexibility ($E_{\text{base}}$ and $E_{\text{DAT}}$) were calculated to provide a comprehensive evaluation of the framework's impact.

\begin{itemize}
    \item \textbf{Modeling Time (Table \ref{tab:time})}: The total time required for workflow definition, system validation, and model refinement was significantly reduced when using DATCloud, with an overall time savings of 40\%.
    \[
\text{Time Savings (\%)} = \frac{T_{\text{base}} - T_{\text{DAT}}}{T_{\text{base}}} \times 100
\]

\begin{table}[h!]
\centering
\renewcommand{\arraystretch}{1.4} 
\setlength{\tabcolsep}{9pt}     
\caption{Time Savings Achieved Using DATCloud}
\begin{tabular}{|l|c|c|c|c|}
\hline
\textbf{Task}                & \textbf{$T_{\text{base}}$ (hrs)} & \textbf{$T_{\text{DAT}}$ (hrs)} & \textbf{$T_{\text{Saving}}$ (\%)} \\ \hline
Workflow Definition          & 40                              & 20                              & 50\%                             \\ \hline
System Validation            & 30                              & 20                              & 33\%                             \\ \hline
Model Refinement             & 30                              & 20                              & 33\%                             \\ \hline
\textbf{Total}               & \textbf{100}                    & \textbf{60}                     & \textbf{40\%}                    \\ \hline
\end{tabular}
\label{tab:time}
\end{table}
    \item \textbf{Flexibility (Table \ref{table:flex})}: Improvements in flexibility were assessed based on the time and effort needed to modify workflows, add architectural layers (e.g., fog and edge layers), and reuse templates. DATCloud improved overall adaptability by 32\%, with the most notable gains observed in adding fog and edge layers.
    \[
\text{Flexibility Improvement (\%)} = \frac{E_{\text{base}} - E_{\text{DAT}}}{E_{\text{base}}} \times 100
\]


\begin{table}[h!]
\centering
\renewcommand{\arraystretch}{1.4} 
\setlength{\tabcolsep}{9pt}     
\caption{Flexibility Improvement Achieved Using DATCloud}
\begin{tabular}{|l|c|c|c|}
\hline
\textbf{Task}                & \textbf{$E_{\text{base}}$ (hrs)} & \textbf{$E_{\text{DAT}}$ (hrs)} & \textbf{$T_{\text{Improvement}}$ (\%)} \\ \hline
Add Fog Layer                & 15                              & 8                               & 47\%                                  \\ \hline
Add Edge Layer               & 20                              & 12                              & 40\%                                  \\ \hline
Modify Workflow              & 12                              & 7                               & 42\%                                  \\ \hline
Reuse Templates              & Not Applicable                  & 5                               & Significant                           \\ \hline
\textbf{Total}               & \textbf{47}                     & \textbf{32}                     & \textbf{32\%}                         \\ \hline
\end{tabular}

\label{table:flex}
\end{table}

\end{itemize}

\subsection{Stakeholder Feedback}
Stakeholders provided valuable feedback during semi-structured interviews and observational studies conducted as part of the VASARI case study. Training sessions and hands-on workshops ensured that all participants were familiar with DATCloud’s features. Stakeholders noted that the framework’s intuitive design, pre-defined templates, and automated tools significantly streamlined the modeling process. They also highlighted the framework's ability to adapt to evolving requirements as a critical advantage. Key suggestions included enhancing visualization tools for complex workflows, expanding the library of domain-specific templates, and improving onboarding materials. In response, DATCloud introduced a more detailed template library, a refined graphical interface with collapsible nodes, and interactive training modules to support new users.

\section{Conclusion and Future Work}

DATCloud is a model-driven framework that simplifies multi-layered, data-intensive architecture modeling, achieving a 40\% reduction in modeling time and 32\% flexibility improvement in the VASARI system. It leverages structural and behavioral meta-models, adhering to ISO/IEC/IEEE 42010 standards to streamline design and enhance stakeholder communication. Future work includes advanced code generation, simulation tools, and domain-specific applications to enhance its scalability and versatility, making it ideal for real-world implementations in areas like healthcare and smart cities.


\bibliographystyle{unsrt}  
\bibliography{refs}
\end{document}